\documentstyle[12pt]{article}

\newcommand{\beq}{\begin{equation}}
\newcommand{\eeq}{\end{equation}}

\newcommand{\pa}{\partial}

\def\rlx{\relax\leavevmode}

\def\IZ{\rlx\hbox{\sf Z\kern-.4em Z}}
\def\IN{\rlx\hbox{\rm I\kern-.18em N}}
\def\IR{\rlx\hbox{\rm I\kern-.18em R}}
\def\IC{\rlx\hbox{\sf C\kern-.5em I }}

\newcommand{\ii}{\mbox{i}}

\newcommand{\dd}{\mbox{d}}
\newcommand{\su}{\mbox{s}}
\newcommand{\cc}{\mbox{c}}

\newcommand{\hD}{\widehat{D}}
\newcommand{\hh}{\widehat{h}}
\newcommand{\hI}{\widehat{I}}
\newcommand{\Hs}{H_{\mbox{s}}}
\newcommand{\Ps}{P_{\mbox{s}}}
\newcommand{\Hc}{H_{\mbox{c}}}
\newcommand{\tilHc}{\widetilde{H}_{\mbox{c}}}
\newcommand{\tilHs}{\widetilde{H}_{\mbox{s}}}
\newcommand{\tilPs}{\widetilde{P}_{\mbox{s}}}

\newcommand{\calJ}{{\cal J}}

\newcommand{\hDel}{\widehat{\Delta}}

\newcommand{\AnS}{{\cal A}_{\mbox{s}}}
\newcommand{\AnC}{{\cal A}_{\mbox{c}}}

\newcommand{\Fs}{{\cal F}_{\mbox{s}}}
\newcommand{\Fc}{{\cal F}_{\mbox{c}}}

\newcommand{\vacS}{|0\rangle_{\mbox{s}}}
\newcommand{\vacC}{|0\rangle_{\mbox{c}}}

\newcommand{\res}{\mbox{Res}}

\title{Common Algebraic Structure for the Calogero-Sutherland Models}
\author{Saburo Kakei\thanks{E-mail: kakei@poisson.ms.u-tokyo.ac.jp}\\[5mm]
{\it Department of Mathematical Sciences, University of Tokyo,}\\
{\it Komaba 3-8-1, Meguro-ku, Tokyo 153, Japan}}
\date{}

\begin{document}
\maketitle

\begin{abstract}
We investigate common algebraic structure for the rational 
and trigonometric Calogero-Sutherland models by using 
the exchange-operator formalism. 
We show that the set of the Jack polynomials whose arguments 
are Dunkl-type operators provides an orthogonal basis for the 
rational case.
\end{abstract}
\bigskip
\bigskip

One dimensional
quantum integrable models with long-range interaction have attracted 
much interest, because of not only their physical significance, but also
their beautiful mathematical structure.
One of such models is the Sutherland (trigonometric) model, 
which describes interacting particles on a circle\cite{Suth}. 
The total momentum and Hamiltonian of the model are 
respectively given by
\beq\begin{array}{rcl}
\Ps & = & \displaystyle
\sum_{j=1}^{N} \frac{1}{\ii}\frac{\pa}{\pa\theta_j},\\
\Hs & = & \displaystyle
      -\sum_{j=1}^{N} \frac{\pa^2}{\pa \theta_j^2}
      + \frac{1}{2}\sum_{j<k}
        \frac{\beta(\beta-1)}{
        \sin^2\left[(\theta_j-\theta_k)/2\right]},
\end{array}
\label{Ham1:Suth}
\eeq
where $\beta$ is a real constant.
Excited states for the Sutherland model are written in terms of the 
Jack symmetric polynomials. 

Another example of long-range interaction is the Calogero
(rational) model of interacting harmonic oscillators\cite{Cal}:
\beq
\Hc = \frac{1}{2}\sum_{j=1}^{N}
      \left( -\frac{\pa^2}{\pa x_j^2} + x_j^2 \right)
      + \sum_{j<k}\frac{\beta(\beta-1)}{(x_j-x_k)^2}.
\label{Ham:Cal}
\eeq
Excited states for this model are of the form, 
$\psi(x) = \varphi(x)\psi^{(\cc)}_0(x)$,
where $\varphi(x)$ is some symmetric polynomial and $\psi^{(\cc)}_0(x)$
is ground state wavefunction (see equation (\ref{eq:gsCal}) below).
The polynomial part $\varphi(x)$ can be obtained in principle
\cite{Suth,Cal}. 
However, properties of orthogonal bases have not been so much
clarified as in the case of the Sutherland model.

Due to the integrability of the Calogero model, the operator 
$H_{\cc}$ belongs to a family of commuting differential operators
({\it ``conserved quantities''} in physical terminology). 
Ujino and Wadati explicitly constructed polynomials that 
diagonalize first two of them\cite{UW1}. 
They further obtained an operator representation for the 
eigenfunctions and showed that they diagonalize
first two of the conserved quantities\cite{UW2}.
Polychronakos studied some special cases of 
the wavefunctions\cite{Poly2} by using the exchange-operator 
formalism\cite{Poly,BHV}.

An orthogonal polynomials associated with the Calogero model were
recently investigated by Baker and Forrester\cite{BF}.
Their proof of orthogonality is based on the orthogonality of 
another set of polynomials, which they call {\it ``generalized 
Jacobi polynomials''}. They obtained the orthogonality of the 
Calogero case via some limiting procedure.
In this letter, using the exchange-operator formalism\cite{Poly,BHV}, 
we shall show that the algebraic structure of the Calogero model 
exactly coincides with that of the Sutherland model.
As a consequence, we present new type of operator representation 
of a basis that diagonalize all of the conserved 
quantities simultaneously,
and prove the orthogonality without taking limiting procedure.

We start with reviewing the method of calculating the excited states 
for the Sutherland model\cite{Suth}.
Let us rewrite the operators (\ref{Ham1:Suth})
in terms of the variables $x_j = \exp(\ii\theta_j)$; then
\begin{eqnarray}
\Ps & = & \sum_{j=1}^{N} x_j\frac{\pa}{\pa x_j},
\label{P2:Suth}\\
\Hs & = & \sum_{j=1}^{N} \left(x_j\frac{\pa}{\pa x_j}\right)^2
      - \beta(\beta-1) \sum_{j<k}
        \frac{2 x_j x_k}{(x_j - x_k)^2}.
\label{Ham2:Suth}
\end{eqnarray}
The ground state wavefunction for the model is
\beq
\psi^{(\su)}_{0}(x) = \prod_{j<k}|x_j - x_k|^{\beta}
               \prod_{j=1}^N x_j^{-\beta(N-1)/2} .
\eeq
To obtain the excited states, it is convenient to make a gauge 
transformation on the momentum and Hamiltonian:
\begin{eqnarray}
\tilPs & = & (\psi^{(\su)}_{0})^{-1} \circ \Ps \circ \psi^{(\su)}_{0}
       \;=\; \sum_{j=1}^N x_j \frac{\pa}{\pa x_j},\label{eP:Suth}\\
\tilHs & = & (\psi^{(\su)}_{0})^{-1} \circ \Hs \circ \psi^{(\su)}_{0}
             - \frac{\beta^2}{12}N(N^2-1) \nonumber\\
       & = & \sum_{j=1}^N 
             \left( x_j\frac{\pa}{\pa x_j} \right)^2
             + \beta \sum_{j<k}\frac{x_j+x_k}{x_j-x_k}
             \left( x_j\frac{\pa}{\pa x_j} - x_k\frac{\pa}{\pa x_k}\right).
             \label{eHam:Suth}
\end{eqnarray}
A basis of joint eigenspace for 
$\tilPs$ and $\tilHs$ are known as the Jack polynomials\cite{Mac,St}.
The Jack polynomials $J_{\lambda}(x)$, indexed by the partitions 
$\lambda=(\lambda_1, \ldots, \lambda_N)$ of length $\leq N$,
are uniquely determined by the following properties:
\begin{enumerate}
\item $\displaystyle J_{\lambda}(x) = 
      m_{\lambda} + \sum_{\mu(<\lambda)}u_{\lambda\mu}m_{\mu}$ ,
\item $J_{\lambda}(x)$ are eigenfunctions of $\tilHs$,
\end{enumerate}
where $m_{\lambda}$ are the monomial symmetric functions, 
and $\mu<\lambda$ is defined by the dominance ordering\cite{Mac}.
Instead of the second property, we may impose the orthogonality 
with respect to the scalar product,
\beq
(f,g)_{\mbox{s}}=
\oint f(x^{-1})g(x) 
\psi^{(\su)}_{0}(x^{-1})\psi^{(\su)}_{0}(x) 
\prod_{j=1}^N\frac{\dd x_j}{2\pi\ii x_j},
\label{firstN}
\eeq
where the integration contour is the unit circle in the complex plane.

Integrability of the Sutherland model, i.e. existence of a family of 
commuting operators that includes the Sutherland Hamiltonian,
 can be proved by using the exchange-operator formalism\cite{Poly}.
We first introduce so-called {\it ``Dunkl operators''}\cite{Dun}:
\beq
D_j = \frac{\pa}{\pa x_j}
      - \beta \sum_{k(\neq j)}\frac{1}{x_j - x_k}(s_{jk}-1)
\qquad (j=1,\ldots,N),
\label{eq:Dunkl}
\eeq
where $s_{ij}$ are elements of the symmetric group $S_N$.
An element $s_{ij}$ acts on functions of $x_1$, $\ldots$, $x_N$ 
as an operator which permutes arguments $x_i$ and $x_j$.
These operators satisfy the following properties:
\begin{eqnarray}
& & [ D_i, D_j] \;=\; [x_i, x_j] \;=\; 0,\nonumber\\
& & s_{ij}D_j \;=\; D_i s_{ij}, \qquad
s_{ij} D_k \;=\; D_k s_{ij} \quad ( k\neq i,j ),\\[0em]
& & [ D_i, x_j ] \;=\; \delta_{ij} 
\left(1+\beta{\displaystyle \sum_{k(\neq i)}}s_{ik}\right)
             -(1-\delta_{ij})\beta s_{ij} .\nonumber
\label{eq:CRdx}
\end{eqnarray}
We denote the algebra generated by the elements $x_j$, $D_j$ and 
$s_{ij}$ as $\AnS$.
We then introduce an $\AnS$-module $\Fs$ ({\it ``Fock space''})
generated by the vacuum vector $\vacS =1$.
The elements $D_j$ of $\AnS$ annihilate the vacuum vector, 
and $s_{ij}$ preserve $\vacS$:
\beq
D_j \vacS = 0,\qquad s_{ij} \vacS = \vacS .
\label{eq:piVac}
\eeq

Further we define {\it ``Cherednik operators''} $\hD_j$
\cite{Ch,BGHP}:
\begin{eqnarray}
\hD_j & = & x_j D_j + \beta \sum_{k(<j)} s_{jk}\nonumber\\
 & = & x_j\frac{\pa}{\pa x_j}
       - \beta \sum_{k(<j)}\frac{x_k}{x_j - x_k} (s_{jk}-1)
       - \beta \sum_{k(>j)}\frac{x_j}{x_j - x_k} (s_{jk}-1)
       + \beta (j-1) .
\end{eqnarray}
Since the operators $\hD_j$ commute each other, they are diagonalized
simultaneously by suitable choice of the bases of $\Fs$
\cite{BGHP,Opdam}.
We introduce {\it ``non-symmetric Jack polynomials''} 
$\calJ^{\lambda}_{w}(x)$ with $w\in S_N$, 
characterized by the following properties 
\cite{BGHP,Opdam}:
\begin{enumerate}
\item $\displaystyle \calJ^{\lambda}_{w}(x) = 
      x^{\lambda}_{w} 
      + \sum_{(\mu,w')<(\lambda,w)}
        C^{\lambda \mu}_{w w'} x^{\mu}_{w'}$ ,
\item $\calJ^{\lambda}_{w}(x)$ are joint eigenfunctions for 
      the operators $\hD_j$,
\end{enumerate}
where we have used the notation 
$x^{\lambda}_{w} = 
x^{\lambda_1}_{w(1)} \cdots x^{\lambda_N}_{w(N)}$.
The ordering $(\mu,w')<(\lambda,w)$ is defined as follows:
\beq
(\mu,w')<(\lambda,w) \quad \Longleftrightarrow \quad
\left\{\begin{array}{ll}
(\mbox{i}) & \mu < \lambda,\\
(\mbox{ii}) & \mbox{if } \mu = \lambda \mbox{ then the first non-vanishing}\\
    & \mbox{difference }w(j)-w'(j)\mbox{ is positive}.
\end{array}\right.
\eeq
For the element $w_0$ of $S_N$
such that $w_0(j)=N-j+1$ ($j=1,\ldots,N$), 
eigenvalues of $\hD_j$ are given by
\beq
\hD_j \calJ^{\lambda}_{w_0}(x) = 
\left\{ \lambda_{N-j+1} + \beta(j-1) \right\}
\calJ^{\lambda}_{w_0}(x).
\label{eq:nonsymJ}
\eeq
For other elements $w\in S_N$, eigenvalues of $\hD_j$ are all obtained
by permutating the components of the multiplet 
$\left\{ \lambda_{N-j+1} + \beta(j-1) \right\}_{j=1,\ldots,N}$.

Using $\hD_j$, we introduce generating function of commuting 
operators\cite{BGHP}:
\beq
\hDel_{\su}(u) = \prod_{j=1}^N (u + \hD_j).
\eeq
If we expand $\hDel_{\su}(u)$ as polynomial in $u$, the coefficients 
$\hI^{(\su)}_j$
form a set of commuting operators.
The transformed momentum (\ref{eP:Suth}) and 
Hamiltonian (\ref{eHam:Suth}) of the Sutherland model are related 
to $\hI^{(\su)}_j$;
\beq\begin{array}{rcl}
\displaystyle\res\: \hI^{(\su)}_1 & = & \displaystyle
\tilPs + \frac{\beta}{2}N(N-1),\\
\displaystyle\res\: \left((\hI^{(\su)}_1)^2-2\hI^{(\su)}_2 \right) 
 & = & \displaystyle
\tilHs + \beta (N-1)\tilPs + \frac{\beta^2}{6}N(N-1)(2N-1).
\end{array}\eeq
where $\res\:X$ means that action of $X$ is restricted to symmetric 
functions of the variables $x_1, \ldots, x_N$.

Since $\hD_{\su}(u)$ is symmetric in $\hD_j$, 
symmetric eigenfunctions are obtained by symmetrizing 
$\calJ^{\lambda}_{w}(x)$, 
i.e. the Jack polynomials $J_{\lambda}(x)$ are the eigenfunctions.
Eigenvalues of $\hDel_{\su}(u)$ are given by
\beq
\hDel_{\su}(u) J_{\lambda}(x) = 
\prod_{j=1}^N
\left\{ u + \lambda_{N-j+1} + \beta(j-1) \right\}
J_{\lambda}(x).
\eeq
Since all the eigenvalues of $\hDel_{\su}(u)$ are distinct and 
the operator $\hD_{\su}(u)$ is self-adjoint with respect to the scalar 
product (\ref{firstN}),
the Jack polynomials $J_{\lambda}(x)$ form an orthogonal basis with 
respect to the scalar product (\ref{firstN}).

We then proceed to the Calogero model.
The ground state for the Calogero Hamiltonian (\ref{Ham:Cal}) is
\beq
\psi^{(\cc)}_0(x) = \prod_{j<k}|x_j - x_k|^{\beta}
               \prod_{j=1}^N \exp\left( -\frac{x_j^2}{2} \right).
\label{eq:gsCal}
\eeq
As in the case of the Sutherland model, the Calogero Hamiltonian is also 
related to the Dunkl operators $D_j$\cite{Poly,BHV}. 
We perform a kind of gauge transformation on $\Hc$:
\begin{eqnarray}
\tilHc & = & 
\prod_{j<k}|x_j - x_k|^{-\beta} \circ \Hc \circ 
\prod_{j<k}|x_j - x_k|^{\beta}
\nonumber\\
 & = & \frac{1}{2}\sum_{j=1}^N\left(-\frac{\pa^2}{\pa x_j^2} + x_j^2\right)
      - \frac{\beta}{2}\sum_{j\neq k}\frac{1}{x_j-x_k}
        \left( \frac{\pa}{\pa x_j} - \frac{\pa}{\pa x_k}\right).
\label{eHam:Cal}
\end{eqnarray}
We then define an analogue of creation and annihilation operators;
\beq
A^{\dag}_j = \frac{1}{\sqrt{2}} (-D_j + x_j), \qquad
A_j = \frac{1}{\sqrt{2}} (D_j + x_j).
\label{eq:CrAnn}
\eeq
We call an algebra generated by $A_j$, $A^{\dag}_j$ and $s_{ij}$
as $\AnC$. 
Since the commutation relations of these operators are the same as
those of $x_j$ and $D_j$,
we can introduce an isomorphism of $\AnS$ to $\AnC$ as follows:
\beq
\rho (x_j) = A^{\dag}_j, \qquad \rho (D_j) = A_j .
\eeq
We note that this kind of isomorphism has already been used in \cite{UW2}.
Here we extend it to the isomorphism of Fock spaces.
Fock space for $\AnC$ is constructed in the same way as
$\AnS$; Fock space $\Fc$ is defined as
$\Fc = \IC [A^{\dag}_1, \ldots, A^{\dag}_N]\vacC$
where the vacuum vector $\vacC = \prod_{j=1}^N \exp(-x_j^2/2)$ is
annihilated by $A^{\dag}_j$, i.e., $A^{\dag}_j\vacC = 0$.
We denote also by $\rho$ the isomorphism of $\Fs$ to $\Fc$ such that
\beq
\rho(\vacS) = \vacC, \qquad
\rho(a |v\rangle) = \rho(a)\rho(|v\rangle)
\label{eq:iso}
\eeq
for $a\in\AnS$ and $|v\rangle\in\Fs$.

Since the operators $\hD_j$ commute each other, we can construct 
commuting operators $\hh_j$ acting on $\Fc$ as
\beq
\hh_j = \rho (\hD_j) = A^{\dag}_j A_j+ \beta \sum_{k(<j)}s_{jk}.
\eeq
Generating function of commuting operators that include $\tilHc$ is 
constructed by using $\hh_j$:
\beq
\hDel_{\cc}(u) = \rho\left(\hDel_{\su}(u)\right)
 = \prod_{j=1}^N (u + \hh_j).
\eeq
We then define $\hI^{(\cc)}_j$ as coefficients of $\hDel_{\cc}(u)$:
\beq
\hDel_{\cc}(u) = \sum_{j=0}^N u^{N-j} \hI^{(\cc)}_j.
\eeq
The transformed Calogero Hamiltonian $\tilHc$ is obtained as 
$\tilHc = \res\:\hI^{(\cc)}_1 + N/2$. 
Our aim is diagonalization of $\hDel_{\cc}(u)$ on $\Fc$. 
Since the Jack polynomials $J_{\lambda}(x)$ ($\in\Fs$) diagonalize 
$\hDel_{\su}(u)$, we conclude that the vectors,
\beq
\rho\left( J_{\lambda}(x)\vacS \right)
 = J_{\lambda}(A^{\dag}) \vacC \;\in\;\Fc ,
\label{eq:OBc}
\eeq
diagonalize $\hDel_{\cc}(u)$.
The eigenvalues of $\hDel_{\cc}(u)$ are the same as those of 
$\hDel_{\su}(u)$ and all of them are distinct.

We then introduce another scalar product,
\begin{eqnarray}
\langle\langle f,g \rangle\rangle 
& = & \,_{\cc}\langle 0| 
   f(A_1,\ldots,A_N)g(A^{\dag}_1,\ldots,A^{\dag}_N) \vacC
\label{eq:SP3}\\
& = & \,_{\su}\langle 0| 
   f(D_1,\ldots,D_N)g(x_1,\ldots,x_N) \vacS
\nonumber
\end{eqnarray}
for $f$ and $g$ homogeneous polynomials of the same degree.
The operator $A^{\dag}_j$ is the adjoint of $A_j$ 
with respect to this scalar product.
Hence $\hDel_{\cc}$ is self-adjoint for (\ref{eq:SP3}).
It follows that the Jack polynomials are pairwise
orthogonal relative to the scalar product (\ref{eq:SP3}).
We note that the second expression of (\ref{eq:SP3}) is 
equivalent to the pairing introduced in \cite{BF} 
(equation (6.4) of \cite{BF}).

It is well-known that the Jack polynomials are pairwise orthogonal 
for two kinds of scalar products; one is (\ref{firstN}) and the other is 
combinatorial one\cite{Mac}.
The scalar product (\ref{eq:SP3}) is the third example.
Recently Polychronakos calculated the norms of the elementary 
symmetric polynomials with respect to this scalar product
\cite{Poly2}. 
The norms for the general Jack polynomials with respect to 
(\ref{eq:SP3}) were evaluated by Baker and Forrester\cite{BF}. 
In our notation, their result is written as
\beq
\langle\langle J_{\lambda}, J_{\mu} \rangle\rangle = 
\delta_{\lambda\mu}
J_{\lambda}(x_1= \cdots = x_N =1)
\prod_{(i,j)\in\lambda}
\left\{\lambda_i -j +1 + \beta (\lambda'_j -i) \right\}.
\label{eq:BF}
\eeq
with $\lambda'=(\lambda'_1,\lambda'_2,\ldots)$ the conjugate partition 
to $\lambda$.
(We remark that the normalization of the Jack polynomials used in 
\cite{BF,St} is different form ours.)
On the other hand, Stanley obtained the following formula
(\cite{St}, Theorem 5.4):
\beq
J_{\lambda}(x_1= \cdots = x_N =1)=
\prod_{(i,j)\in\lambda}
\frac{j-1+\beta(N-i+1)}{\lambda_i -j + \beta(\lambda'_j -i+1)},
\eeq
Combining these results, we finally come to the expression,
\beq
\langle\langle J_{\lambda}, J_{\mu} \rangle\rangle = 
\delta_{\lambda\mu}
\prod_{(i,j)\in\lambda}
\frac{\{j-1+\beta(N-i+1)\}
      \{\lambda_i -j +1 + \beta (\lambda'_j -i)\}}{
      \lambda_i -j + \beta(\lambda'_j -i+1)}.
\label{eq:3rdNorm}
\eeq

In conclusion, we have proved that the algebraic structure 
for the Calogero model exactly coincides with that of 
the Sutherland model by using the exchange-operator formalism.
We further proved that the vectors (\ref{eq:OBc}) $\in\Fc$
form an orthogonal basis for the Calogero model.
We hope that our results provide a useful tool for deeper understanding
of the Calogero model.

\section*{Acknowledgments}
The author acknowledges Dr. Yusuke Kato for directing his
attention to the Calogero-Sutherland models and for 
fruitful discussions.
He also acknowledges Professors Junkichi Satsuma and Tetsuji
Tokihiro for their critical reading of the manuscript and helpful
comments.
Thanks are also due to Professors Katsuhisa Mimachi and 
Masatoshi Noumi for informations on some important references.

\bigskip

\begin{small}
{\it Note added in proof.}
After submission of this letter, a paper \cite{UW3} has been 
brought to the author's attention.
In \cite{UW3}, Ujino and Wadati proved that the vectors (\ref{eq:OBc})
diagonalize the first two of the conserved quantities.
However, they have not proved the orthogonality.
We remark that the Rodrigues-type formula of \cite{UW2,UW3}
is a consequence of the formula in \cite{LV} and the isomorphism 
(\ref{eq:iso}) of the Fock spaces as is suggested in \cite{UW2}.
The author acknowledges Dr. Hideaki Ujino for informing his results
and for helpful comments.
\end{small}


\bigskip

\begin{thebibliography}{16}

\bibitem{Suth}
B. Sutherland:
Phys. Rev. {\bf A4} (1971) 2019-2021;
Phys. Rev. {\bf A5} (1972) 1372-1376.

\bibitem{Cal}
F. Calogero: J. Math. Phys. {\bf 12} (1971) 419-436.

\bibitem{UW1}
H. Ujino and M. Wadati:
J. Phys. Soc. Jpn. {\bf 64} (1995) 2703-2706.

\bibitem{UW2}
H. Ujino and M. Wadati:
J. Phys. Soc. Jpn. {\bf 65} (1996) 653-656.

\bibitem{Poly2}
A.P. Polychronakos: 
{\it ``Quasihole Wavefunctions for the Calogero Model''},
preprint (comd-mat/9603132).

\bibitem{Poly}
A.P. Polychronakos: 
Phys. Rev. Lett. {\bf 69} (1992) 703-705.

\bibitem{BHV}
L. Brink, T.H. Hansson and M. Vasiliev:
Phys. Lett. {\bf B286} (1992) 109-111.

\bibitem{BF}
T.H. Baker and P.J. Forrester:
{\it ``The Calogero-Sutherland Model and Generalized 
Classical Polynomials''},
preprint (solv-int/9608004).

\bibitem{Mac}
I.G. Macdonald:
{\it ``Symmetric Functions and Hall Polynomials''}, Second Edition,
Oxford Mathematical Monographs, Clarendon Press, Oxford, 1995.

\bibitem{St}
R.P. Stanley:
Adv. in Math. {\bf 77} (1988) 76-115.

\bibitem{Dun}
C.F. Dunkl: 
Trans. Amer. Math. Soc. {\bf 311} (1989) 167-183.

\bibitem{Ch}
I. Cherednik:
Invent. Math. {\bf 106} (1991) 411-32.

\bibitem{BGHP}
D. Bernard, M. Gaudin, F.D.M. Haldane and V. Pasquier: 
J. Phys. {\bf A26} (1993) 5219-5236.

\bibitem{Opdam}
E.M. Opdam:
Acta Math. {\bf 175} (1995) 75-121.

\bibitem{UW3}
H. Ujino and M. Wadati:
J. Phys. Soc. Jpn. {\bf 65} (1996) 2423-39.

\bibitem{LV}
L. Lapointe and L. Vinet:
Commun. Math. Phys. {\bf 178} (1996) 425-52.

\end{thebibliography}
\end{document}